\documentclass[10pt]{iopart}
\usepackage{iopams}
\usepackage{graphicx}
\usepackage{cite}
\newcommand{\text}[1]{\mathrm{#1}}
\newcommand{\eqref}[1]{(\ref{#1})}

\newcommand{\rot}{\mbox{\rm{rot\,}}}
\newcommand{\diverg}{\mbox{\rm{div\,}}}

\begin{document}

\title[Force on a bunch near plasma-vacuum boundary]{Force exerted on particle bunch propagating near plasma-vacuum boundary}

\author{K.V. Lotov}
\address{Budker Institute of Nuclear Physics, Novosibirsk, 630090, Russia}
\address{Novosibirsk State University, Novosibirsk, 630090, Russia}
\ead{K.V.Lotov@inp.nsk.su}

\vspace{10pt}
\begin{indented}
\item[]\today
\end{indented}

\begin{abstract}
If a charged particle bunch propagates near a plasma-vacuum boundary, it excites a surface wave and experiences a force caused by the boundary. For the linearly responding plasma and ultra-relativistic bunch, the spatial distribution of excited fields is calculated, and the force exerted on a short and narrow bunch is approximated by elementary functions. The force attracts the bunch to the boundary, if the bunch is in the vacuum, and repels otherwise. There are also additional focusing and defocusing components of the force.
\end{abstract}

% Uncomment for PACS numbers
%\pacs{00.00, 20.00, 42.10}
\vspace{2pc}
\noindent{\it Keywords}: plasma wakefield acceleration, surface wave, wakefield potential, force

\ioptwocol
\section{Introduction}

In studies of novel accelerators \cite{RMP81-1229,PoP14-055501}, the setup sometimes appears when an ultra-relativistic charged particle bunch propagates parallel or at a shallow angle to a vacuum-plasma boundary \cite{PRST-AB4-091301,NJP18-103013,PPCF61-104004}, either in the plasma or in the vacuum. The electromagnetic fields induced by the boundary can deflect \cite{PRST-AB4-091301}, attract \cite{NJP18-103013}, or destroy \cite{PPCF61-104004} the bunch, so the boundary effect is of practical importance. In this paper, we derive analytical expressions for these fields under the assumption of linear plasma response and approximate them by elementary functions wherever possible.

The solution method is similar to that used for calculating Cherenkov radiation of a point charge moving along the boundary between two media \cite{JAP26-2,UFN4-781}. There is, however, an important difference. In our case, the field harmonics cannot freely propagate in media and decay away from the boundary.

In section~\ref{s2}, we describe the solution method in general and obtain amplitudes of Fourier harmonics for fields of a point charge. These amplitudes are valid for an arbitrary speed of the charge. In section~\ref{s3}, we restrict our consideration to ultra-relativistic charges, plasma and vacuum as adjoining media, and wakefield potential as a convenient characteristic of the forces exerted on moving beams. We analyze the potential distribution behind the point charge in the cases of large and small charge-to-boundary distances. In section~\ref{s4}, we calculate the fields in the vicinity of the charge and present approximate expressions for them. In section~\ref{s5}, we calculate the force exerted on a short and narrow Gaussian beam.

\section{Fields of a point charge, general formalism}
\label{s2}

\begin{figure}[tb]
\centering\includegraphics{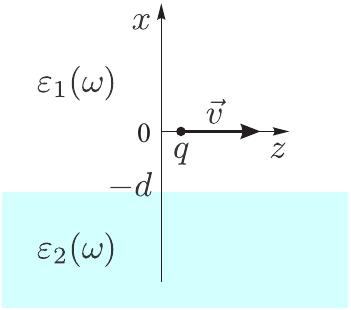}
\caption{Geometry of the problem.}\label{fig1-geometry}
\end{figure}

We consider the point charge $q$ moving with the velocity $v$ in the $z$-direction parallel to the plane boundary between two media with relative permittivities $\varepsilon_1 (\omega)$ and $\varepsilon_2 (\omega)$, respectively. We assume $\varepsilon_i (\omega) \leq 1$ for $i=1,2$. The charge is in the first medium at $x=0$, and the boundary is at $x=-d$ (figure~\ref{fig1-geometry}). The charge and current densities of the point charge are
\begin{equation}\label{e1}
    \rho_b (\vec{r}, t) = q \delta (\vec{r} - \vec{v} t), \ \
    \vec{j}_b (\vec{r}, t) = \vec{v} \rho_b, \ \
    \vec{v} = (0, 0, v),
\end{equation}
and their Fourier transforms are
\begin{eqnarray}
    \rho_\omega (\vec{k}, \omega) &=& \frac{1}{(2 \pi)^2} \int \rho_b (\vec{r},t) e^{-i \vec{k} \vec{r} + i \omega t} d\vec{r} dt \nonumber
\\ \label{e2-2}
    &=& \frac{q}{2 \pi} \delta (\omega - k_z v),
\\ \label{e2-3}
    \vec{j}_\omega (\vec{k}, \omega) &=& \frac{q\vec{v}}{2 \pi} \delta (\omega - k_z v).
\end{eqnarray}

The solution for fields consists of three parts: fields of the charge in a whole space filled with medium "1" (here called the incoming wave and labelled by the subscript ``0''), fields in the upper half-space with no external charge (reflected wave, subscript ``1''), and fields in the lower half-space (transmitted wave, subscript ``2''). The sum of fields ``0''and ``1'' must match fields ``2'' at the boundary. The electromagnetic potential $(\varphi, \vec{A})$ for the incoming wave is conveniently written in the Lorentz gauge, which reads as
\begin{equation}\label{e3}
    \diverg \vec{A}_0 + \frac{\varepsilon_1}{c} \frac{\partial \varphi_0}{\partial t} = 0
\end{equation}
for any frequency harmonic; here $c$ is the speed of light. The potentials for the secondary waves are in the Coulomb gauge:
\begin{equation}\label{e4}
    \diverg \vec{A}_1 = \diverg \vec{A}_2 = 0, \qquad
    \varphi_1 = \varphi_2 = 0.
\end{equation}
The fields are
\begin{equation}\label{e5}
    \vec{E} = - \nabla \varphi - \frac{1}{c} \frac{\partial \vec{A}}{\partial t}, \qquad
    \vec{B} = \rot \vec{A}.
\end{equation}

For any frequency harmonic, the potentials of the incoming wave obey the equations
\begin{eqnarray}\label{e6-7}
     \triangle \varphi_0 - \frac{\varepsilon_1}{c^2} \frac{\partial^2 \varphi_0}{\partial t^2} = - \frac{4 \pi}{\varepsilon_1} \rho_b,
\\ \label{e6-8}
    \triangle \vec{A}_0 - \frac{\varepsilon_1}{c^2} \frac{\partial^2 \vec{A}_0}{\partial t^2} = - \frac{4 \pi}{c} \vec{j}_b,
\end{eqnarray}
where $\triangle$ is the Laplacian, whence
\begin{eqnarray}\label{e7}
    \varphi_0 (\vec{k}, \omega) = \frac{4 \pi \rho_\omega (\vec{k}, \omega)}{\varepsilon_1 (k^2 - \varepsilon_1 \omega^2 / c^2)} = \frac{2 q \delta (\omega - k_z v)}{\varepsilon_1 (k^2 - \varepsilon_1 \omega^2 / c^2)}, \\
\label{e8}
    A_{0z} (\vec{k}, \omega) = \frac{v \varepsilon_1}{c}\varphi_0 (\vec{k}, \omega), \qquad
    A_{0x} = A_{0y} = 0.
\end{eqnarray}
To find field harmonics at the boundary, we must integrate over $k_x$. We also integrate over $k_z$ to shorten the result:
\begin{eqnarray} \nonumber
    \varphi_0 (x, k_y, z, \omega) =
    \frac{1}{2 \pi} \int \varphi_0 (\vec{k}, \omega) e^{i k_x x + i k_z z} dk_x dk_z = \\ \label{e9}
    = \frac{q}{\pi v \varepsilon_1} \int \frac{e^{i k_x x + i\omega z/v}}{k_x^2 + \kappa_1^2} dk_x
    = \frac{q e^{-\kappa_1 |x| + i\omega z/v}}{v \varepsilon_1 \kappa_1},
\end{eqnarray}
where
\begin{equation}\label{e10}
    \kappa_i = \sqrt{k_y^2 + \omega^2/v^2 - \varepsilon_i \omega^2 / c^2}
\end{equation}
is real and positive for $\varepsilon_i \leq 1$. Then the potentials are
\begin{eqnarray}\label{e11-13}
    \varphi_0 (\vec{r}, t) = \frac{q}{2 \pi v} \int \frac{ dk_y d \omega}{\varepsilon_1 \kappa_1} e^{-\kappa_1 |x| + i k_y y + i\omega (z/v - t)},
\\ \label{e11-14}
    A_{0z} (\vec{r}, t) = \frac{q}{2 \pi c} \int \frac{ dk_y d \omega}{\kappa_1} e^{-\kappa_1 |x| + i k_y y + i\omega (z/v - t)}.
\end{eqnarray}

In what follows, we use three types of quantities distinguished by their arguments: the quantities in real space [e.g., $E_{0x} (\vec{r}, t)$], harmonic amplitudes [$E_{0x} (x)$] and harmonic amplitudes at the boundary ($E_{0x}$). From equations \eqref{e5}, \eqref{e11-13} and \eqref{e11-14}, we find
\begin{eqnarray}\label{e13a-15}
    E_{0x} (\vec{r}, t) = \int E_{0x} (x) e^{i k_y y + i\omega (z/v - t)} dk_y d \omega,
\\ \label{e13a-16}
    E_{0x} (x) = \pm\frac{q e^{-\kappa_1 |x|}}{2 \pi v \varepsilon_1} = \mp E_{0x} e^{-\kappa_1 |x| + \kappa_1 d},
\\ \label{e13b-17}
    E_{0x} = -\frac{q e^{-\kappa_1 d}}{2 \pi v \varepsilon_1},
\end{eqnarray}
where upper (lower) signs are for $x>0$ ($x<0$). Other components of the electric field are
\begin{eqnarray}\label{e13c}
    E_{0y} (x) = -\frac{i q k_y e^{-\kappa_1 |x|}}{2 \pi v \varepsilon_1 \kappa_1}, \qquad
    E_{0y} = \frac{i k_y}{\kappa_1} E_{0x},
\\ \label{e13d-19}
    E_{0z} (x) = -\frac{i q \omega e^{-\kappa_1 |x|}}{2 \pi v^2 \varepsilon_1 \kappa_1} \left( 1 - \frac{v^2 \varepsilon_1}{c^2} \right),
\\ \label{e13d-20}
    E_{0z} = \frac{i \omega}{v \kappa_1} \left( 1 - \frac{v^2 \varepsilon_1}{c^2} \right) E_{0x}.
\end{eqnarray}

Similarly, vector potential harmonics for reflected and transmitted waves obey
\begin{equation}\label{e12}
    \triangle \vec{A}_i - \frac{\varepsilon_i}{c^2} \frac{\partial^2 \vec{A}_i}{\partial t^2}
    = \frac{\partial^2 \vec{A}_i}{\partial x^2} - \kappa_i^2 \vec{A}_i = 0,
\end{equation}
whence the solutions localized near the boundary are
\begin{equation}\label{e13}
    \vec{A}_{i} (\vec{r}, t) = \int \vec{A}_i e^{-\kappa_i |x+d| + i k_y y + i\omega (z/v - t)} dk_y d \omega.
\end{equation}

At the boundary, tangential components of electric and magnetic fields must be continuous. Denoting
\begin{equation}\label{e21-23}
    k_{0x} = -i \kappa_1, \ \
    k_{1x} = i \kappa_1, \ \
    k_{2x} = -i \kappa_2, \ \
    k_z = \omega/v,
\end{equation}
we can write the magnetic field in the usual form
\begin{equation}\label{e20-24}
    \vec{B}_i = \frac{c}{\omega} \left[ \vec{k}_i \times \vec{E}_i \right],
\end{equation}
whence the boundary conditions follow:
\begin{eqnarray}\label{e22-25}
    E_{0y} + E_{1y} = E_{2y}, \qquad
    E_{0z} + E_{1z} = E_{2z},
\\ \nonumber
    k_{0x} E_{0y} - k_y E_{0x} + k_{1x} E_{1y} - k_y E_{1x} = k_{2x} E_{2y} - k_y E_{2x}, \\ \label{e22-26}
\\ \nonumber
    k_z E_{0x} - k_{0x} E_{0z} + k_z E_{1x} - k_{1x} E_{1z} = k_z E_{2x} - k_{2x} E_{2z}.
\\ \label{e23-27}
\end{eqnarray}
To complete the system, we also need
\begin{equation}\label{e24}
    \diverg \vec{E}_i = 0, \qquad
    k_{ix} E_{ix} + k_y E_{iy} + k_z E_{iz} = 0
\end{equation}
and equation
\begin{equation}\label{e25}
    \varepsilon_1 (E_{0x} + E_{1x}) = \varepsilon_2 E_{2x},
\end{equation}
which is redundant, but helpful. Then we simplify \eqref{e22-26}-\eqref{e23-27} with the help of \eqref{e22-25}, \eqref{e25}, and \eqref{e21-23}:
\begin{eqnarray} \nonumber
    \kappa_1 (E_{0y} - E_{1y}) = \kappa_2 (E_{0y} + E_{1y}) + i k_y (\varepsilon_2/\varepsilon_1 - 1) E_{2x}, \\ \label{e26}
\\ \nonumber
    \kappa_1 (E_{0z} - E_{1z}) = \kappa_2 (E_{0z} + E_{1z}) + i k_z (\varepsilon_2/\varepsilon_1 - 1) E_{2x},
\\ \label{e27}
\end{eqnarray}
combine these equations to exclude $E_{1y}$ and $E_{1z}$ with the help of \eqref{e24}:
\begin{eqnarray} \nonumber
    (\kappa_1 - \kappa_2) (-k_{0x} E_{0x}) = (\kappa_2 + \kappa_1) \bigl(-k_{1x} (E_{2x} \varepsilon_2/\varepsilon_1 - E_{0x})\bigr) \\ \label{e29}
    + i (\varepsilon_2/\varepsilon_1 - 1) (k_y^2 + k_z^2) E_{2x},
\end{eqnarray}
and take into account \eqref{e10} to obtain
\begin{equation}\label{e33}
    E_{2x} = \frac{2 \varepsilon_1 \kappa_1}{\varepsilon_2 \kappa_1 + \varepsilon_1 \kappa_2} E_{0x}.
\end{equation}
From \eqref{e25} and \eqref{e33}, we find
\begin{equation}\label{e34}
    E_{1x} = \frac{\varepsilon_2 \kappa_1 - \varepsilon_1 \kappa_2}{\varepsilon_2 \kappa_1 + \varepsilon_1 \kappa_2} E_{0x}.
\end{equation}
From \eqref{e26}, \eqref{e33}, and \eqref{e13c},
\begin{equation}\label{e36}
    E_{1y} = i k_y \left( \frac{ 2 \varepsilon_1}{\varepsilon_2 \kappa_1 + \varepsilon_1 \kappa_2} - \frac{1}{\kappa_1} \right) E_{0x}.
\end{equation}
From \eqref{e24}, \eqref{e34}, and \eqref{e36},
\begin{equation}\label{e38}
    E_{1z} = -\frac{i v}{\omega} \left( \frac{2 \varepsilon_1 k_y^2 + \kappa_1 (\varepsilon_2 \kappa_1 - \varepsilon_1 \kappa_2)}{\varepsilon_2 \kappa_1 + \varepsilon_1 \kappa_2} - \frac{k_y^2}{\kappa_1} \right) E_{0x}.
\end{equation}
From \eqref{e22-25}, \eqref{e13c}, and \eqref{e36},
\begin{equation}\label{e40}
    E_{2y} = \frac{2 i k_y \varepsilon_1}{\varepsilon_2 \kappa_1 + \varepsilon_1 \kappa_2} E_{0x}.
\end{equation}
From \eqref{e24}, \eqref{e33}, and \eqref{e40},
\begin{equation}\label{e42}
    E_{2z} = \frac{2i v \varepsilon_1 (\kappa_2 \kappa_1 - k_y^2)}{\omega (\varepsilon_2 \kappa_1 + \varepsilon_1 \kappa_2)} E_{0x}.
\end{equation}
Expressions \eqref{e33}--\eqref{e42} have the form
\begin{equation}\label{e-39}
    E_{i\alpha} = a_{i\alpha} E_{0x},
\end{equation}
with all diversity of field components hidden in coefficients $a_{i\alpha}$. The same coefficients determine vector potentials and electric fields of the secondary waves ($i=1,2$):
\begin{eqnarray} \nonumber
    \vec{A}_i (\vec{r}, t) = \frac{i q c}{2 \pi v} \int \frac{\vec{a}_i e^{-\kappa_i |x+d| -\kappa_1 d + i k_y y + i\omega (z/v - t)}}{\omega \varepsilon_1} dk_y d \omega,
\\ \label{e-40}
\\ \nonumber
    \vec{E}_i (\vec{r}, t) = - \frac{q}{2 \pi v} \int \frac{\vec{a}_i e^{-\kappa_i |x+d| -\kappa_1 d + i k_y y + i\omega (z/v - t)}}{\varepsilon_1} dk_y d \omega.
\\ \label{e-41}
\end{eqnarray}
Note that putting
\begin{equation}\label{e44-42}
    g_1 = i \kappa_1, \qquad
    g_2 = - i \kappa_2, \qquad
    \mu_i = 1
\end{equation}
in formulae (III.131)--(III.135) of Ref.~\cite{UFN4-781} does not produce the same result, so the difference between Cherenkov radiation and the considered problem goes beyond a simple change of variables.

\section{Wakefield potentials}
\label{s3}

Now we restrict our consideration to ultra-relativistic charges ($v=c$) and the plasma-vacuum boundary. In the plasma,
\begin{equation}\label{e-43}
    \varepsilon_i = 1-\omega_p^2/\omega^2 \equiv \varepsilon, \qquad
    \kappa_i = \sqrt{k_y^2 + \omega_p^2/c^2} \equiv \kappa,
\end{equation}
where $\omega_p$ is the plasma frequency. It is convenient to characterize the fields by the wakefield potential $\Phi = \varphi - A_z$, the gradient of which determines the force exerted on an axially moving (in the direction $\vec{e}_z$) unit charge:
\begin{equation}\label{e66a-44}
    \vec{E} + \left[ \vec{e}_z \times \vec{B} \right] = -\nabla \Phi, \qquad
    \Phi (x) = \frac{i E_z (x)}{k_z}.
\end{equation}

There are two regimes of interest (the charge moving in vacuum or in plasma) and two regions in each, which we distinguish with capital subscripts:
\begin{displaymath}
\begin{array}{llllll}
  % after \\: \hline or \cline{col1-col2} \cline{col3-col4} ...
  A: & \text{vacuum}, & \varepsilon_1 = 1, & \kappa_1 = |k_y|, & \Phi_A = \Phi_0 + \Phi_1; \\
  \hline
  B: & \text{plasma}, & \varepsilon_2 = \varepsilon, & \kappa_2 = \kappa, & \Phi_B = \Phi_2; \\[2mm]
  C: & \text{plasma}, & \varepsilon_1 = \varepsilon, & \kappa_1 = \kappa, & \Phi_C = \Phi_0 + \Phi_1; \\
  \hline
  D: & \text{vacuum}, & \varepsilon_2 = 1, & \kappa_2 = |k_y|, & \Phi_D = \Phi_2. \\
\end{array}
\end{displaymath}

From equations \eqref{e66a-44}, \eqref{e13d-19}, \eqref{e38}, \eqref{e42}, and \eqref{e13b-17}, we obtain
\begin{eqnarray}\label{e54}
    \Phi_A (x) = \frac{q c |k_y| (\kappa - |k_y|)}{\pi \omega^2 (\kappa + \varepsilon |k_y|)} e^{-|k_y| (x+2d)},
\\ \label{e55}
    \Phi_B (x) = \frac{q c |k_y| (\kappa - |k_y|)}{\pi \omega^2 (\kappa + \varepsilon |k_y|)} e^{-|k_y| d + \kappa (x+d)}.
\end{eqnarray}
Similarly,
\begin{eqnarray}\nonumber
    \Phi_C (x) &= \frac{q (1 - \varepsilon)}{2 \pi c \varepsilon \kappa} \left( e^{-\kappa |x|} - e^{-\kappa (x+2d)} \right)
\\ \label{e56}
    &+ \frac{q c |k_y| (\kappa - |k_y|)}{\pi \omega^2 (\kappa + \varepsilon |k_y|)} e^{-\kappa (x+2d)},
\\ \label{e57}
    \Phi_D (x) &= \frac{q c |k_y| (\kappa - |k_y|)}{\pi \omega^2 (\kappa + \varepsilon |k_y|)} e^{-\kappa d + |k_y| (x+d)}.
\end{eqnarray}
The wakefield potential is continuous at the boundary:
\begin{equation}\label{e58}
    \Phi_A (-d) = \Phi_B (-d), \qquad \Phi_C (-d) = \Phi_D (-d).
\end{equation}

\begin{figure*}[t]
\centering\includegraphics{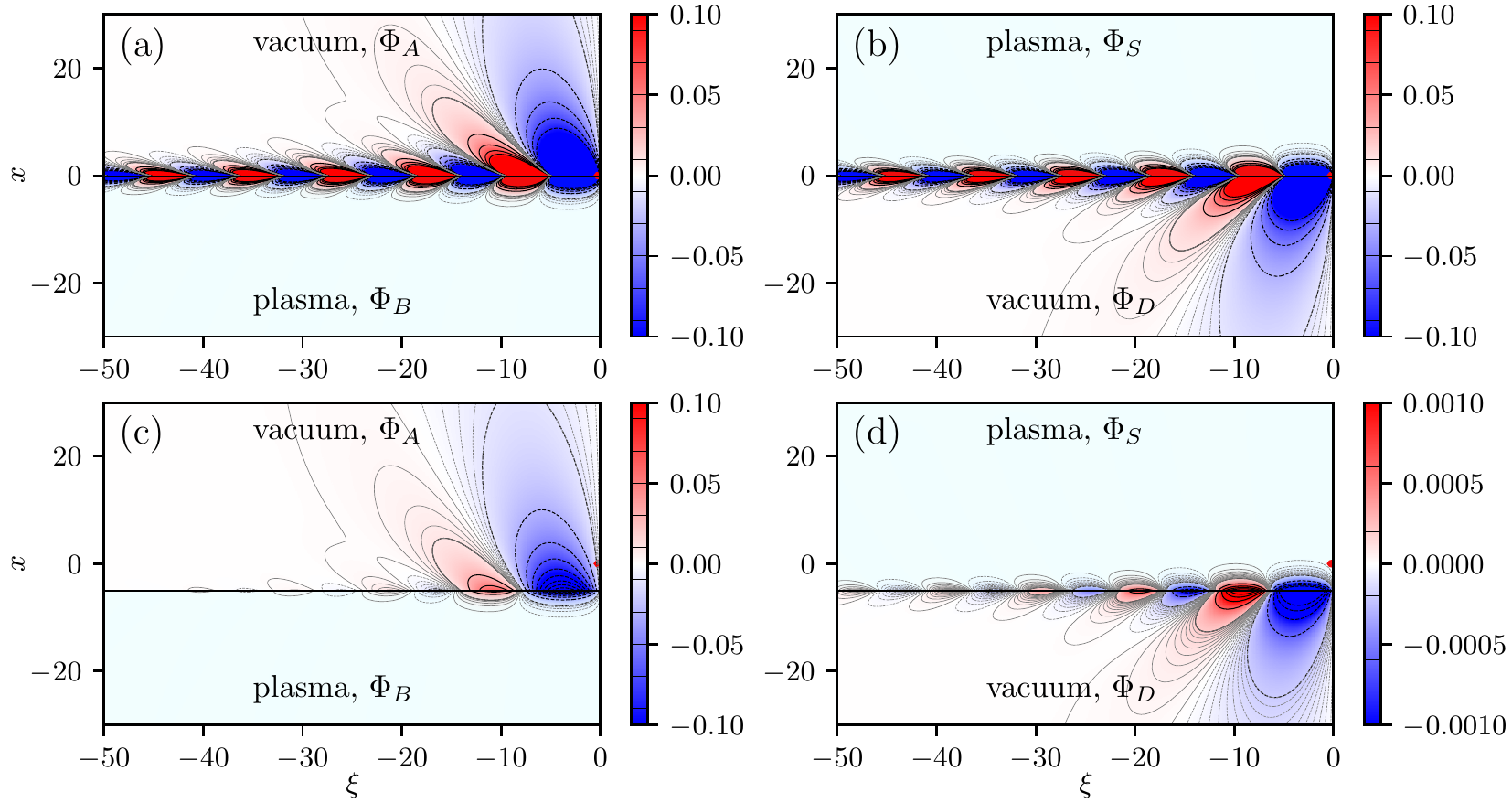}
\caption{The wakefield potential of the surface wave in the plane $y=0$ for the charge propagating in vacuum (a),(c) and in plasma (b),(d) at small [$d=0.1$] (a),(b) and large [$d=5$] (c),(d) distances to the boundary. Small red circles show the charge locations. The color palette is chosen to emphasize small potential values.}\label{fig2-F(x,xi)}
\end{figure*}

From now on, it is more convenient to use dimensionless quantities. We measure frequencies in $\omega_p$, distances in $c/\omega_p$, and potentials in $q \omega_p/c$. Also introduce
\begin{equation}\label{e63-50}
    \omega_0^2 = \frac{|k_y|}{|k_y| + \kappa}, \quad
    \Phi_a = \frac{\omega_0^2 (\kappa - |k_y|)}{\pi(\omega^2 - \omega_0^2)}, \quad
    \xi = z-t,
\end{equation}
and omit arguments $(\vec{r}, t)$ of the dimensionless potentials. The potentials take the form
\begin{eqnarray}\label{e59}
    \Phi_A &= \int \Phi_a e^{-|k_y| (x+2d) + i k_y y + i\omega \xi} dk_y d \omega,
\\ \label{e60}
    \Phi_B &= \int \Phi_a e^{-|k_y| d + \kappa (x+d) + i k_y y + i\omega \xi} dk_y d \omega,
\\ \nonumber
    \Phi_C &= \int \frac{e^{i k_y y + i\omega \xi}}{2 \pi \kappa (\omega^2 - 1)} \left( e^{-\kappa |x|} - e^{-\kappa (x+2d)} \right) dk_y d \omega
\\ \label{e61}
    &+ \int \Phi_a e^{-\kappa (x+2d) + i k_y y + i\omega \xi} dk_y d \omega,
\\ \label{e62}
    \Phi_D &= \int \Phi_a e^{-\kappa d + |k_y| (x+d) + i k_y y + i\omega \xi} dk_y d \omega.
\end{eqnarray}
When integrating \eqref{e59}--\eqref{e62} over $\omega$, we bypass singularities from above to ensure casuality:
\begin{equation}\label{e65-55}
    \int_{-\infty}^\infty \frac{e^{i\omega \xi}}{\omega^2 - \omega_0^2} d \omega =
    \left\{\begin{array}{ll}
    2 \pi \sin (\omega_0 \xi) / \omega_0, & \xi < 0, \\
    0, & \xi \geq 0.
    \end{array} \right.
\end{equation}
Then behind the charge (at $\xi < 0$),
\begin{eqnarray} \nonumber
    \Phi_A &= \int 2 \omega_0 (\kappa - |k_y|) \sin (\omega_0 \xi) e^{-|k_y| (x+2d) + i k_y y} dk_y,
\\ \label{e66}
\\ \nonumber
    \Phi_B &= \int 2 \omega_0 (\kappa - |k_y|) \sin (\omega_0 \xi) e^{-|k_y| d + \kappa (x+d) + i k_y y} dk_y,
\\ \label{e67}
\\ \nonumber
    \Phi_C &= \sin \xi \int \frac{e^{i k_y y}}{\kappa} \left( e^{-\kappa |x|} - e^{-\kappa (x+2d)} \right) dk_y
\\ \label{e68}
    &+ \int 2 \omega_0 (\kappa - |k_y|) \sin (\omega_0 \xi) e^{-\kappa (x+2d) + i k_y y} dk_y,
\\ \nonumber
    \Phi_D &= \int 2 \omega_0 (\kappa - |k_y|) \sin (\omega_0 \xi) e^{-\kappa d + |k_y| (x+d) + i k_y y} dk_y.
\\ \label{e69}
\end{eqnarray}
The first integral in \eqref{e68} can be taken analytically. If
\begin{eqnarray}\label{e70-60}
    |x| = r \cos\phi, \quad y = r \sin\phi, \quad r = \sqrt{x^2 + y^2},
\\ \label{e70-61}
    k_y = \sinh\alpha, \quad \kappa = \sqrt{k_y^2 + 1} = \cosh\alpha,
\end{eqnarray}
then \cite{AS}
\begin{eqnarray} \nonumber
    \int_{-\infty}^\infty \frac{e^{-\kappa |x| + i k_y y} }{\kappa} d k_y
    = \int_{-\infty}^\infty e^{-r (\cosh\alpha \cos\phi - i \sinh\alpha \sin\phi)} d \alpha =
\\ \label{e71}
    = \int_{-\infty}^\infty e^{-r \cosh (\alpha - i\phi)} d (\alpha - i\phi) = 2 K_0 (r),
\end{eqnarray}
whence
\begin{eqnarray}\label{e72-63}
    &\Phi_C = \Phi_W + \Phi_R + \Phi_S,
\\ \label{e72-64}
    &\Phi_W = 2 K_0 \left( \sqrt{x^2 + y^2} \right) \sin \xi,
\\ \label{e72-65}
    &\Phi_R = - 2 K_0 \left( \sqrt{(x+2d)^2 + y^2} \right) \sin \xi,
\\ \label{e72-66}
    &\Phi_S = \int 2 \omega_0 (\kappa - |k_y|) \sin (\omega_0 \xi) e^{-\kappa (x+2d) + i k_y y} dk_y.
\end{eqnarray}
We refer to these three terms as regular wakefield ($\Phi_W$), reflection ($\Phi_R$), and surface ($\Phi_S$) contributions. The regular wakefield term is not related to the boundary and is well known in the theory of plasma-based accelerators \cite{PAcc20-171,PAcc22-81,PF30-252}. The reflection term is the wakefield of the image charge of the opposite sign.

\begin{figure}[tb]
\centering\includegraphics{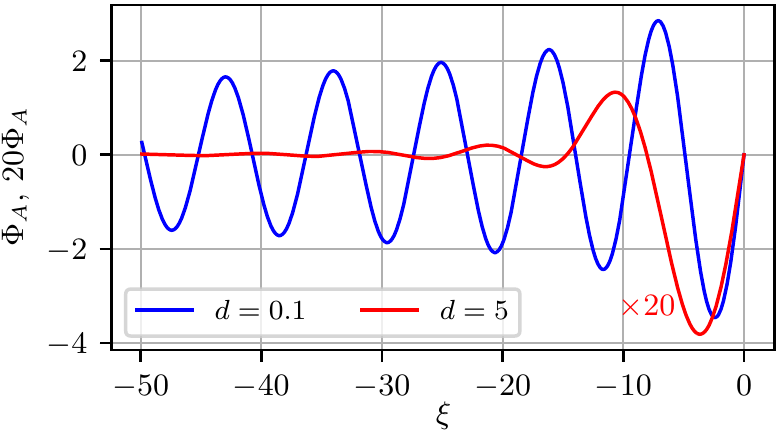}
\caption{Behavior of the wakefield potential $\Phi_A$ behind the charge (at $x=y=0$) for small ($d=0.1$) and large ($d=5$) distances to the boundary. The graph for $d=5$ is stretched 20 times in the vertical direction. }\label{fig3-typical}
\end{figure}
\begin{figure}[tb]
\centering\includegraphics{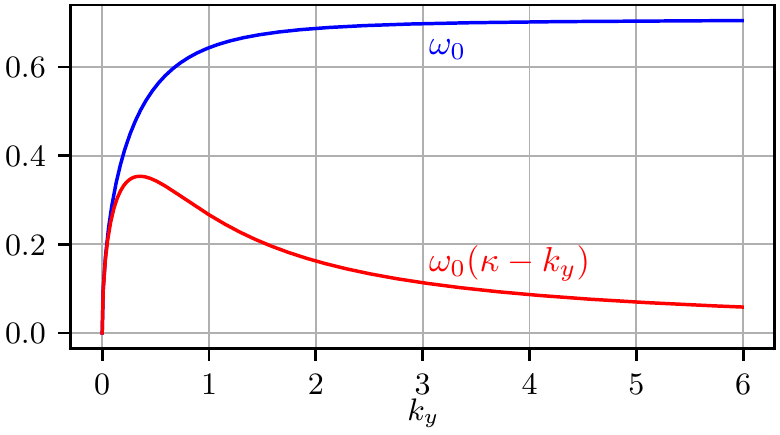}
\caption{The functions contained in integrands of the wakefield potential.}\label{fig4-functions}
\end{figure}

The surface term $\Phi_S$ and potentials $\Phi_A$, $\Phi_B$, and $\Phi_D$ describes the surface waves or surface plasmon polaritons. Properties of individual Fourier harmonics of these waves are well known \cite{LL8,Raether}. However, we are interested in properties of harmonic superposition in the particular case of wave excitation by an ultra-relativistic point charge. Expressions \eqref{e66}, \eqref{e67}, \eqref{e69}, and \eqref{e72-66} cannot be reduced to elementary or well-known special functions, so we analyze them numerically. Their behavior qualitatively differ in cases of large and small charge-to-boundary distances (figure~\ref{fig2-F(x,xi)}). If the charge is close to the boundary ($d \lesssim 1$), then the wave lasts a long time behind the charge and oscillates with the period $2\pi/\sqrt{2}$ (figure~\ref{fig3-typical}). This is because the oscillating factors $\sin (\omega_0 \xi)$ in the integrands have approximately the same frequency $\omega_0 \approx 1/\sqrt{2}$ at $k_y \gg 1$ (figure~\ref{fig4-functions}). Contributions of different $k_y$ coherently add and form a long-lasting wave, if they are not suppressed by the exponential factors, which is possible for $d \lesssim 1$. The wave also extends transversely, covering a large area [figure~\ref{fig5-F(y,xi)}(a)].

Otherwise ($d \gg 1$), the coherence of contributing harmonics is quickly lost, and the wave exactly behind the charge disappears after a couple of oscillations [figure~\ref{fig2-F(x,xi)}(c),(d)]. The wave energy propagates transversely along the boundary in the form of longitudinally localized wave packets [figure~\ref{fig5-F(y,xi)}(b)].

\begin{figure}[tb]
\centering\includegraphics{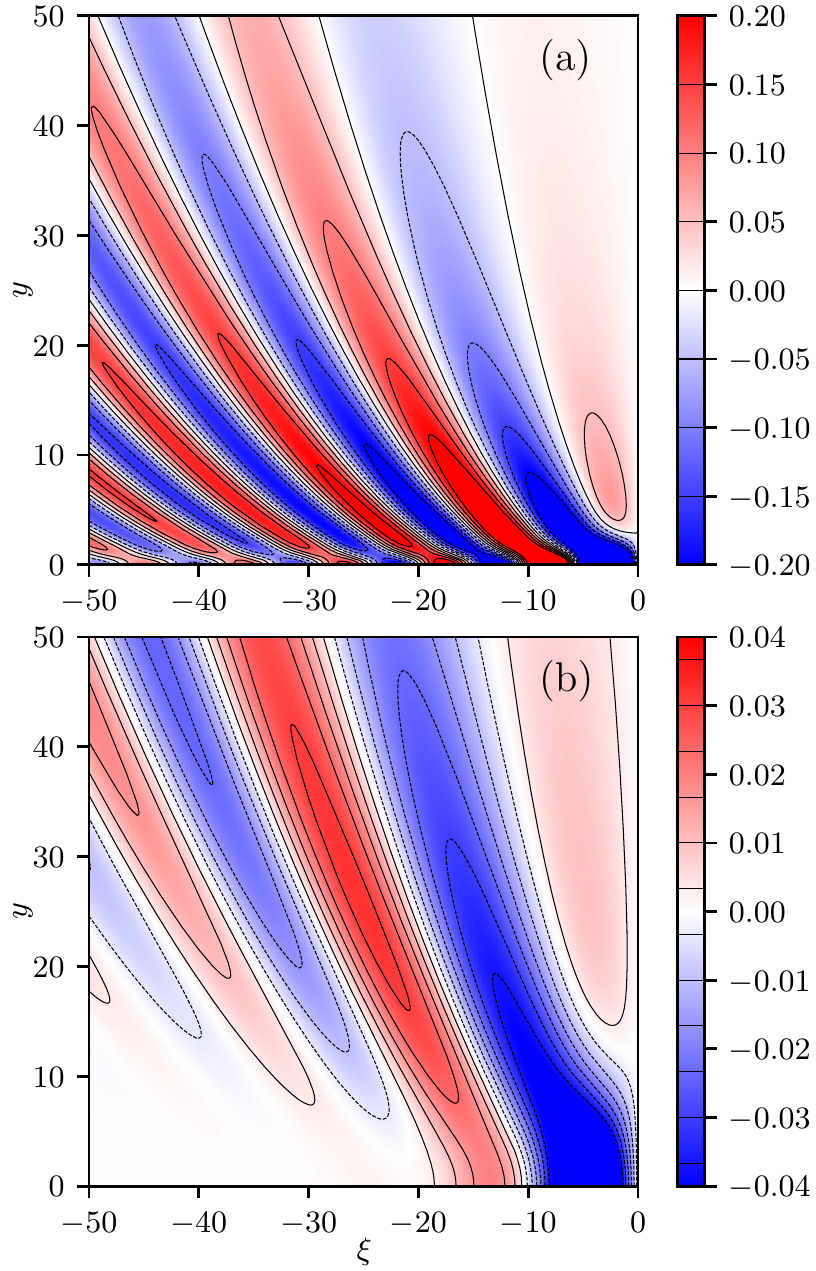}
\caption{The wakefield potential of the surface wave in vacuum in the plane $x=0.9$ for $d=0.1$ (a) and in the plane $x=0$ for $d=5$ (b). The color palette is chosen to emphasize small potential values.}\label{fig5-F(y,xi)}
\end{figure}

In all cases, the potential reaches its maximum values at the surface and, as we will show in section~\ref{s4}, decreases according to a power law in the vacuum and faster than exponentially in the plasma (figure~\ref{fig2-F(x,xi)}). The charge propagation axis ($x=y=0$) defines the symmetry plane $y=0$, but does not hold a distinguished position in this plane.

\begin{figure}[tb]
\centering\includegraphics{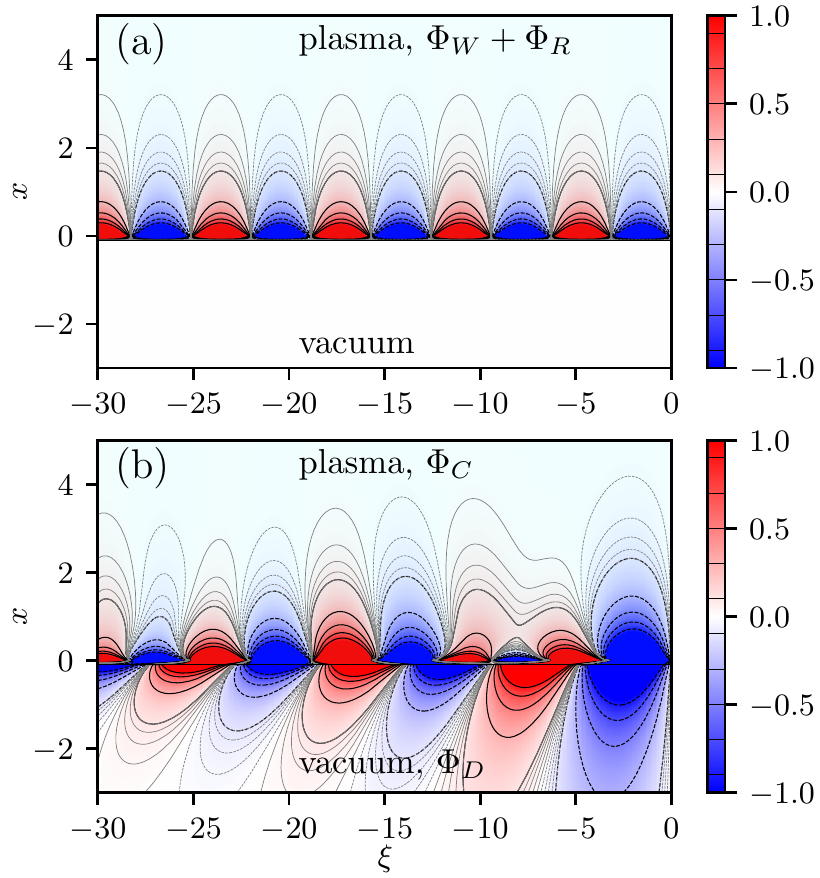}
\caption{The wakefield terms (a) and the total wakefield potential (b) in the plane $y=0$ for the charge propagating in the plasma at the distance $d=0.1$ from the surface. The color palette is chosen to emphasize small potential values.}\label{fig6-wakefield}
\end{figure}

The wakefield terms $\Phi_W$ and $\Phi_R$ that appear if the charge propagates in the plasma, qualitatively differ from the surface wave [figure~\ref{fig6-wakefield}(a)]. They are strictly periodic in $\xi$, and the field energy remains near the charge propagation axis. Unlike the surface wave amplitude, which grows, the sum $\Phi_W + \Phi_R$ decreases as the charge ($d \to 0$) or observation point ($x \to -d$) approaches the boundary. If both surface and wakefield waves are strong, their interference can form exotic field patterns [figure~\ref{fig6-wakefield}(b)]

\section{Fields near the charge}
\label{s4}

Here we calculate the fields in the vicinity of the charge and analyze approximate expressions for them. Since the wakefield terms in \eqref{e72-63} can be straightforwardly differentiated, we focus on approximations for the surface wave only.

\begin{figure}[tb]
\centering\includegraphics{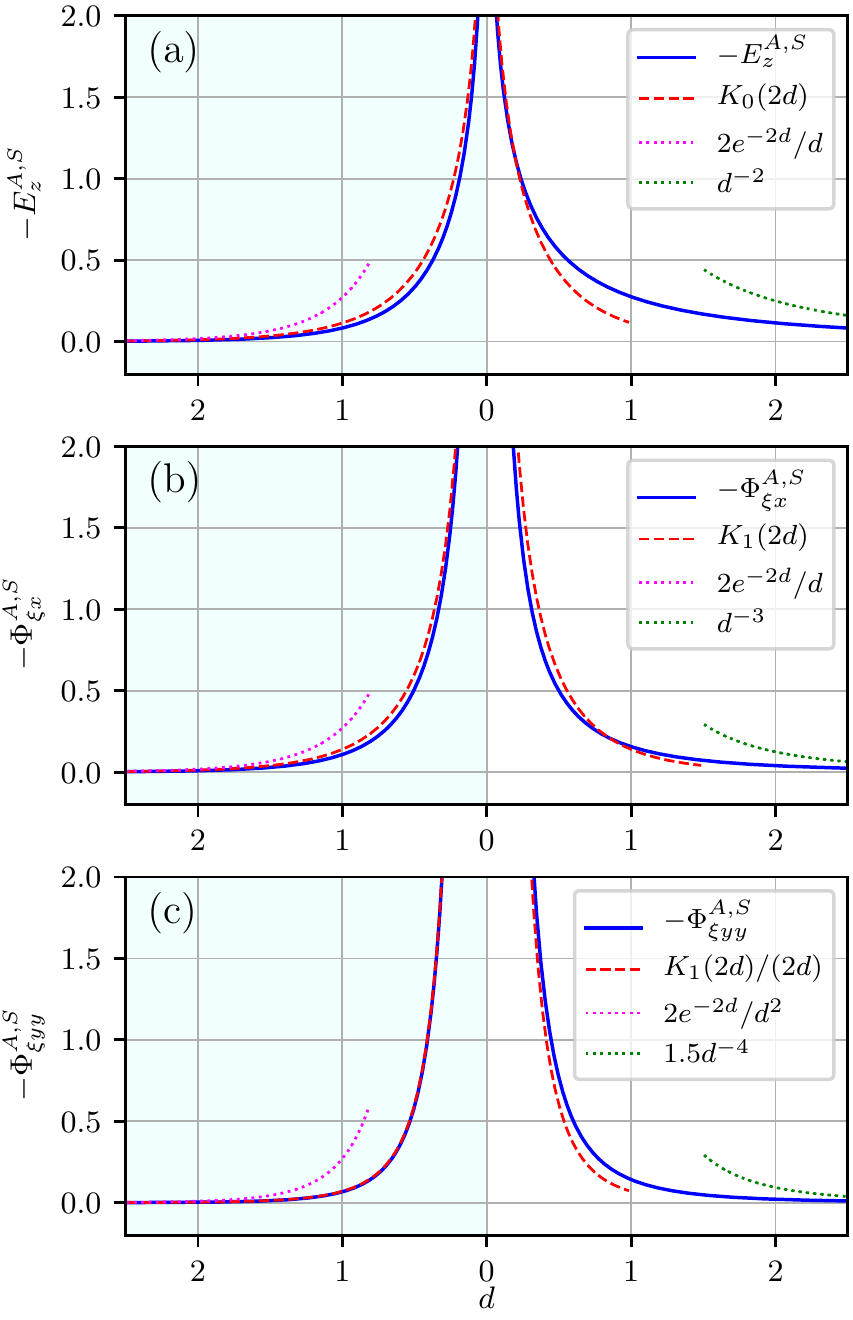}
\caption{The coefficients defining the force components in $z$ (a), $x$ (b), and $y$ (c) directions and their approximations versus charge-to-surface distance $d$ in the plasma (left parts) and in the vacuum (right parts).}\label{fig7-Fs}
\end{figure}

The longitudinal field $E_z^{A,S}$ of the surface wave tends to a constant near the charge:
\begin{equation}\label{e-67}
    E_z^{A,S} = -\left.\frac{\partial \Phi}{\partial \xi}\right|_{x=y=\xi=0} = \int_0^\infty \frac{4 k_y (k_y - \kappa)}{\kappa + k_y} e^{-2 d s} dk_y,
\end{equation}
where $s=k_y$ in the vacuum ($\Phi = \Phi_A$), and $s=\kappa$ in the plasma ($\Phi = \Phi_S$). The field $E_z$ at the surface (at $x=-d$, $y=0$, $\xi=-0$) is expressed by the same formula, but with $d$ replaced by $d/2$. Since the longitudinal field at charge location determines the total energy deposited to the wave, the dependencies $E_z^{A,S} (d)$ also show how the wave amplitude scales with~$d$ [figure~\ref{fig7-Fs}(a)]. The longitudinal field of the reflection wave is positive,
\begin{equation}\label{e-68}
    E_z^R = -\left.\frac{\partial \Phi_R}{\partial \xi}\right|_{x=y=\xi=0} = 2 K_0 (2d),
\end{equation}
so this wave reduces the energy deposited to the wakefield. The longitudinal component of the regular wakefield diverges at $x=y=0$, and its calculation requires a finite drive bunch size to be taken into account.

The force $F_x$ exerted in $x$-direction on a moving unit charge is proportional to $\xi$, and its strength is characterized by the following derivatives:
\begin{eqnarray} \nonumber
    F_x = E_x - B_y \approx -\xi \left.\frac{\partial^2 \Phi}{\partial \xi \partial x}\right|_{x=y=\xi=0} \\ \label{e-69}
    - \xi x \left.\frac{\partial^3 \Phi}{\partial \xi \partial^2 x}\right|_{x=y=\xi=0}
    \equiv -\xi (\Phi_{\xi x}'' + x \Phi_{\xi xx}''').
\end{eqnarray}
For the surface wave, the leading term is
\begin{equation}\label{e-70}
    \Phi_{\xi x}'' = \Phi_{\xi x}^{A,S} \equiv \int_0^\infty \frac{4 k_y s (k_y - \kappa)}{\kappa + k_y} e^{-2 d s} dk_y
\end{equation}
[figure~\ref{fig7-Fs}(b)]. It is negative, so the force attracts a charge of the same sign (as $q$) to the surface. For the reflection wave,
\begin{equation}\label{e-71}
    \Phi_{\xi x}'' = \Phi_{\xi x}^R = 2 K_1 (2d),
\end{equation}
and this force repels a like charge from the surface. The wakefield force formally diverges at $x=y=0$, but  equals zero if the drive beam has a finite size and axisymmetric charge distribution.

The force $F_y$ is symmetric with respect to the $y=0$ plane:
\begin{equation}\label{e-72}
    F_y = E_y + B_x \approx -\xi y \left.\frac{\partial^3 \Phi}{\partial \xi \partial^2 y}\right|_{x=y=\xi=0} \equiv -\xi y \Phi_{\xi yy}'''.
\end{equation}
Its quantitative measures are
\begin{equation}\label{e-73}
    \Phi_{\xi yy}''' = \Phi_{\xi yy}^{A,S} \equiv \int_0^\infty \frac{4 k_y^3 (k_y - \kappa)}{\kappa + k_y} e^{-2 d s} dk_y < 0
\end{equation}
for the surface wave [figure~\ref{fig7-Fs}(c)], and
\begin{equation}\label{e-74}
    \Phi_{\xi yy}''' = \Phi_{\xi yy}^R = \frac{K_1 (2d)}{d} > 0
\end{equation}
for the reflection wave. In the plasma, both are small compared to the focusing force of the regular wakefield
\begin{equation}\label{e-75}
    F_y^W = 2 K_1 (y) \sin \xi.
\end{equation}
In the vacuum,
\begin{equation}\label{e-76a}
    \Phi_{\xi yy}''' = - \Phi_{\xi xx}''' < 0.
\end{equation}

\begin{figure}[tb]
\centering\includegraphics{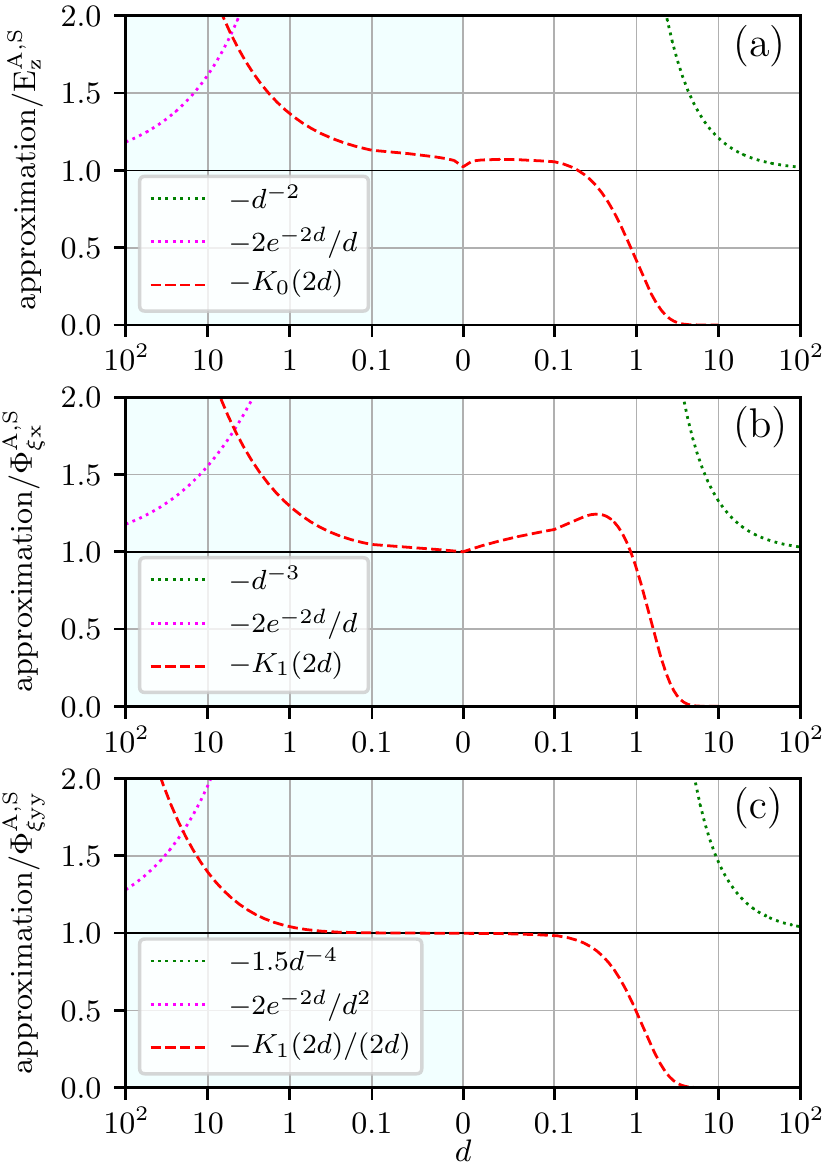}
\caption{The accuracy of force approximations in $z$ (a), $x$ (b), and $y$ (c) directions versus charge-to-surface distance $d$ in the plasma (left parts) and in the vacuum (right parts).}\label{fig8-dFs}
\end{figure}

The quantities \eqref{e-67}, \eqref{e-70}, and \eqref{e-73} characterizing the surface wave force components have the form
\begin{equation}\label{e-76}
    F(d) = \int_0^\infty \frac{4 k_y^m \kappa^n (k_y - \kappa)}{\kappa + k_y} e^{-2 d s} dk_y.
\end{equation}
Because of the exponential term, for $d \gg 1$, the major contribution to \eqref{e-76} comes from $k_y \ll 1$. Putting $\kappa \approx 1$ in this region, we obtain in the vacuum
\begin{equation}\label{e-77}
    F(d) \approx -\int_0^\infty 4 k_y^m e^{-2 d k_y} dk_y = -\frac{m!}{2^{m-1} d^{m+1}}.
\end{equation}
In the plasma, we must keep the second expansion term in the exponent:
\begin{eqnarray}\nonumber
    F(d) &\approx -\int_0^\infty 4 k_y^m e^{-2 d - k_y^2 d} dk_y \\ \label{e-78}
    &= -\frac{2 e^{-2d} ((m-1)/2)!}{d^{(m+1)/2}}.
\end{eqnarray}
However, expressions \eqref{e-77}--\eqref{e-78} slowly approach the approximated functions and can be used only for $d \gg 10$ (figure~\ref{fig8-dFs}, dotted lines).

To approximate $F(d)$ at $d \ll 1$, we make the change of variables \eqref{e70-61}. In the plasma,
\begin{eqnarray} \nonumber
    F(d) = - \int_0^\infty (\sinh \alpha)^{m-1} (\cosh \alpha)^n \left( 1 - e^{-4 \alpha} \right) \\ \label{e-79}
    \times e^{-2 d \cosh \alpha} d \alpha.
\end{eqnarray}
The major contribution to this integral is made at the interval of $\alpha$ where $2 \cosh \alpha \sim e^\alpha \sim 1/d$. At this interval, $e^{-4 \alpha} \sim d^4 \ll 1$. Neglecting $e^{-4 \alpha}$ in \eqref{e-79}, we obtain \cite{AS}
\begin{eqnarray}
    \label{e-80} m=1, \ n=0, \qquad F(d) &\approx -K_0(2d), \\
    \label{e-81} m=1, \ n=1, \qquad F(d) &\approx -K_1(2d), \\
    \label{e-82} m=3, \ n=0, \qquad F(d) &\approx -K_1(2d)/(2d).
\end{eqnarray}
These approximations are rather accurate up to $d \sim 1$ (figures~\ref{fig7-Fs},\,\ref{fig8-dFs}). In the vacuum, we additionally neglect the difference $e^{-\alpha}$ between $\cosh \alpha$ and $\sinh \alpha$ and obtain the same approximations \eqref{e-80}-\eqref{e-82}.

\begin{figure}[tb]
\centering\includegraphics{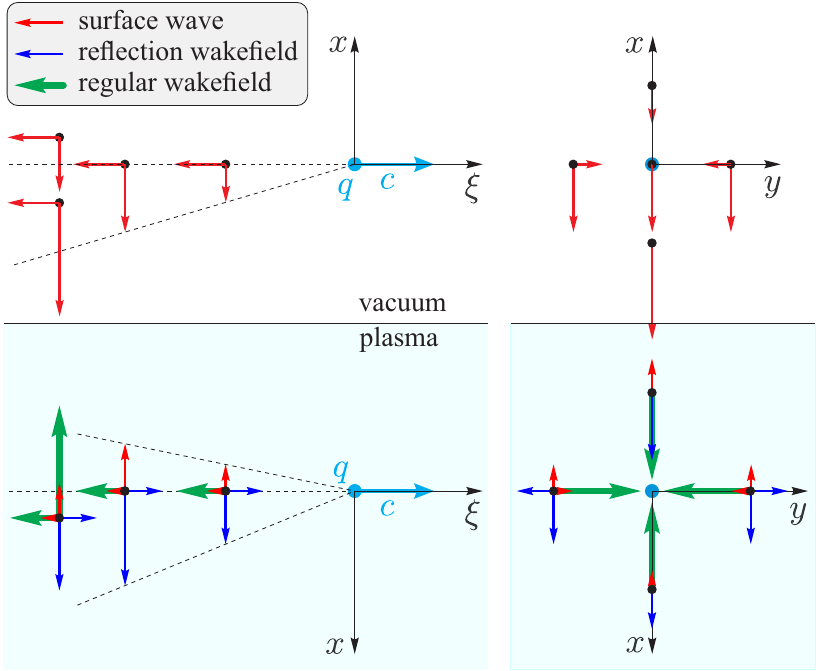}
\caption{Schematic representation of forces acting on a witness test charges in the vacuum (top) and in the plasma (bottom), side (left) and front (right) views.}\label{fig9-forces}
\end{figure}

Figure~\ref{fig9-forces} summarizes the force map in the vicinity of the drive charge $q$. The point-like driver itself experiences no transverse forces and is only decelerated. In the vacuum, the test charge (witness) of the same sign is attracted to the surface with the force \eqref{e-69} that grows linearly with the distance from the driver. The closer the witness to the surface, the stronger the force, according to \eqref{e-76a}. If the witness is not in the $y=0$ plane, it additionally experiences a focusing force \eqref{e-72}. In the plasma, the witness is repelled from the surface, as the repelling force of the reflection wakefield \eqref{e-71} is approximately twice stronger than the attractive force of the surface wave \eqref{e-70}, \eqref{e-81}. If the witness is not on the driver propagation axis, it is focused by the regular wakefield that dominates other two forces. The longitudinal forces are approximately constant in the considered area. The forces of surface wave and regular wakefield decelerate the witness, while the reflection wakefield accelerates it.

\section{Force exerted on a particle bunch}
\label{s5}

Here we calculate the force exerted on different parts of a Gaussian particle bunch with the charge density
\begin{equation}\label{e1a-84}
    \rho_b = \frac{Q}{(2\pi)^{3/2} \sigma_x \sigma_y \sigma_z} \exp \left( -\frac{x^2}{2 \sigma_x^2} - \frac{y^2}{2 \sigma_y^2} - \frac{\xi^2}{2 \sigma_z^2} \right)
\end{equation}
assuming the bunch is small:
\begin{equation}\label{e-85}
    \sigma_x \ll c/\omega_p, \qquad  \sigma_y \ll c/\omega_p, \qquad  \sigma_z \ll c/\omega_p.
\end{equation}
We focus on force components caused by the boundary and will not write out the force of the regular wakefield, as the latter has been extensively studied and allows for analytical expressions for the on-axis longitudinal field only \cite{PoP12-063101}.

\begin{figure}[tb]
\centering\includegraphics{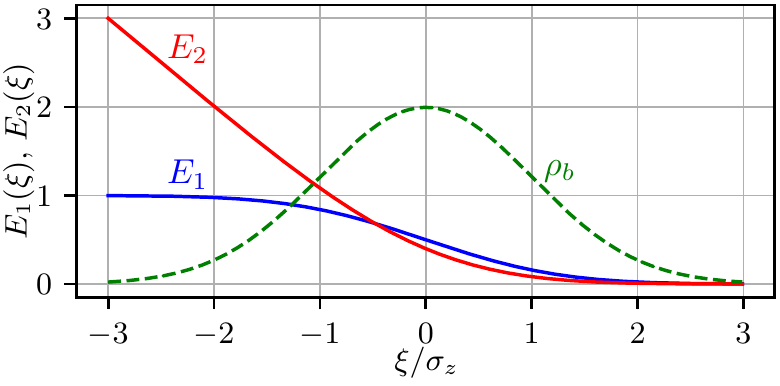}
\caption{Functions $E_1 (\xi)$ and $E_2 (\xi)$ that determine longitudinal distributions of force components. The dashed line shows the beam density profile. }\label{fig10-Es}
\end{figure}

The surface wave exerts a force on a test unit charge
\begin{eqnarray} \nonumber
    \vec{F}_t (x,y,\xi) = \int_\xi^\infty d\xi' \int dx' \int dy' \\ \label{e88-86}
    \times \vec{F}(x-x', y-y', \xi-\xi') \rho_b (x',y',\xi'),
\end{eqnarray}
where $\vec{F} (x,y,\xi)$ is the force created by a unit driver charge located at $x=y=\xi=0$. For a small bunch, we can approximate this force with expansions obtained in the previous section. Assume, for definiteness, that the bunch propagates in the vacuum, and introduce $\vec{\rho} = \vec{r}-\vec{r}'$. Then
\begin{eqnarray} \nonumber
    \left\{ F_{tx}, F_{ty}, F_{tz} \right\} = Q \int_{-\infty}^0 d\rho_\xi \int_{-\infty}^\infty d\rho_x \int_{-\infty}^\infty d\rho_y
\\ \nonumber
    \times \frac{\left\{  -\rho_\xi \left(\Phi_{\xi x}^A + \rho_x \Phi_{\xi xx}^A \right), \  -\rho_\xi \rho_y \Phi_{\xi yy}^A, \ E_z^A \right\}}{(2\pi)^{3/2} \sigma_x \sigma_y \sigma_z}
\\ \nonumber
    \times \exp \left( -\frac{(x-\rho_x)^2}{2 \sigma_x^2} - \frac{(y-\rho_y)^2}{2 \sigma_y^2} - \frac{(\xi-\rho_\xi)^2}{2 \sigma_z^2} \right)
\\ \nonumber
    = Q \left\{  E_2 (\xi) \left(\Phi_{\xi x}^A + x \Phi_{\xi xx}^A \right), \  y E_2 (\xi) \Phi_{\xi yy}^A, \ E_1(\xi) E_z^A \right\},
\\ \label{e-87}
\end{eqnarray}
where we denote (figure~\ref{fig10-Es})
\begin{eqnarray}
\label{e-88}
  E_1 (\xi) &= \frac{1}{2}\text{erfc} \left( \frac{\xi}{\sigma_z \sqrt{2}} \right)
  =  \frac{1}{\sqrt{\pi}} \int_{\xi/(\sigma_z \sqrt{2})}^\infty e^{-t^2} dt,
\\ \label{e-89}
  E_2 (\xi) &= - \frac{\xi}{2}\text{erfc} \left( \frac{\xi}{\sigma_z \sqrt{2}} \right) + \frac{\sigma_z}{\sqrt{2\pi}} e^{-\xi^2 / (2 \sigma_z^2)}.
\end{eqnarray}
The forces of surface wave and reflection wakefield in the plasma differ only in superscripts ($S$ or $R$ instead of $A$). To restore the dimensions, the expression \eqref{e-87} must be additionally multiplied by $\omega_p^2 / c^2$ while keeping the expressions in braces dimensionless.

\ack

This work was supported by the Russian Fund for Basic Research, project 19-02-00243.

\end{document}